# POSSIBILITIES OF LIFE AROUND ALPHA CENTAURI B

POSIBILIDADES DE VIDA ALREDEDOR DE ALFA CENTAURO B

A. González[a ‡], R. Cárdenas-Ortiz[a †] and J. Hearnshaw[b *]

a) Departamento de Física, Universidad Central "Marta Abreu" de Las Villas, Cuba. agnoa@uclv.cu[‡]; rcardenas@uclv.edu.cu[†]
b) Department of Astronomy, University of Canterbury, New Zealand. john.hearnshaw@canterbury.ac.nz*
† corresponding author

We make a preliminary assessment on the habitability of potential rocky exoplanets around Alpha Centauri B. We use several indexes: the Earth Similarity Index, a mathematical model for photosynthesis, and a biological productivity model. Considering the atmospheres of the exoplanets similar to current Earth's atmosphere, we find consistent predictions of both the Earth Similarity Index and the biological productivity model. The mathematical model for photosynthesis clearly failed because does not consider the temperature explicitly. For the case of Alpha Centauri B, several simulation runs give 11 planets in the habitable zone. Applying to them above mentioned indexes, we select the five exoplanets more prone for photosynthetic life; showing that two of them in principle have better conditions than Earth for this kind of life.

Se hace una estimación preliminar de la habitabilidad de potenciales exoplanetas rocosos en el sistema Alfa del Centauro B. Se usan varios índices de habitabilidad: el Índice de Similaridad con la Tierra, un modelo matemático de fotosíntesis, y un modelo de productividad biológica. Considerando las atmósferas de los exoplanetas similares a la actual de la Tierra, se encuentran predicciones consistentes del Índice de Similaridad con la Tierra y el modelo de productividad biológica. El modelo matemático de fotosíntesis falló, al no considerar la temperatura explícitamente. Para el sistema Alfa del Centauro B, varias simulaciones computacionales predicen la potencial formación de 11 exoplanetas en la zona habitable. Aplicando los mencionados índices, se seleccionan los cinco exoplanetas con mayor potencial para la vida fotosintética; mostrándose que dos de ellos en principio tienen mejores condiciones para este tipo de vida que la Tierra actual.

## INTRODUCTION

The stellar system Alpha Centauri is nowadays one of the main targets in the search for life in our cosmic neighborhood, as it is the closest to the Solar System and observations indicate the absence of gaseous giant planets (potential absorbers of the rocky ones).

On the other hand, computer simulations show the possibility of the existence of several Earth-like planets in the habitable zone of Alpha Centauri B [1]. In this paper we present a preliminary assessment concerning the habitability of exoplanets predicted in the above reference.

## MATERIALS AND METHODS

*Planetary Models*　　We use the results of computer simulations presented in [1]. The number and characteristics of exoplanets depend on the quantity of protoplanets assumed at the start of the simulations. In above mentioned paper, authors present the results of 8 computer simulations, which give as a result the formation of 21 planets. For the characteristics of the predicted planets, we recommend the reader to check [1].

Of the above mentioned 21 planets, we considered in our calculations only those 11 that are inside the zone of habitability of the star (according to the standard opinion of the presence of liquid water in the planetary surface).

*Habitability indexes*　　We use three indexes to assess the possibilities for life: the Earth Similarity Index, the P model for biological productivity, and the E model for photosynthesis.

The Earth Similarity Index (*ESI*) was developed in the Habitability Lab of the University of Puerto Rico at Arecibo [2]. It is defined as

$$ESI = \prod_{i=1}^{n}\left(1-\left|\frac{x_i - x_{i0}}{x_i + x_{i0}}\right|\right)^{\frac{w_i}{n}}, \qquad (1)$$

where $x_i$ is a planetary property, $x_{i0}$ is its value on Earth today, $w_i$ is a weight exponent and $n$ is the number of planetary properties. The weights of the properties considered are summarized in Table I below.

Table I
Environmental variables considered in the Earth Similarity Index

| Planetary property | Reference value | Weight exponent |
| --- | --- | --- |
| Surface temperature | 288 K | 5,58 |
| Escape velocity | 1,0 E. u. (Earth units) | 0,70 |
| Mean radius | 1,0 E. u. | 0,57 |
| Bulk density | 1,0 E. u. | 1,07 |

Not all above properties are independent, but actually the set is designed to consider two environments of the planet: outer and inner. Surface temperature and escape velocity account for the speed of biochemical reactions at planet's surface and for retention of the atmosphere (outer ESI), while mean radius and bulk density account for geodynamics (inner ESI).

Earth-like planets can be defined as any planetary body with a similar terrestrial composition and a temperate atmosphere. This means that the planet is rocky in composition (silicates) and has an atmosphere suitable for most terrestrial vegetation, including complex life.

Current Earth gives ESI = 1. Planets with ESI in the range 0,8-1 are considered capable of hosting life more or less similar to current Earth's. Those having this index in the range 0,6-0,8 (as present Mars), are either to cold or too hot, and therefore could host extremophiles. A planetary body with ESI < 0.6 is considered not habitable.

The Π model for biological productivity [3] is:

$$\frac{\Pi}{\Pi_{max}} = \left(1 - \left(\frac{T_{opt} - T_S}{T_{opt} - 273}\right)^2\right) f(p_{CO_2}), \quad (2)$$

where $\Pi$ is the biological productivity, $\Pi_{max}$ is the maximum possible bio-productivity, $T_{opt}$ is the optimum temperature for life, $T_S$ is the temperature at planet's surface and $f(p_{CO_2})$ is a function of the carbon dioxide pressure in the atmosphere (because this gas is needed for photosynthesis). In this work we are using a similar model atmosphere for all exoplanets: current Earth's. Therefore, the numerical value for $f(p_{CO_2})$ is assumed to be the same for all them. It is obtained assuming $T_{opt}$ = 298 K and average current Earth's conditions: $T_S$ = 288 K and $\Pi/\Pi_{max}$=0,5 [3].

The E model for photosynthesis is:

$$\frac{P}{P_S}(z) = \frac{1 - \exp[-E_{PAR}(z)/E_S]}{1 + E^*_{UV}(z)}, \quad (3)$$

where $P(z)$ is the phytoplankton photosynthesis rate at depth z in the ocean, $P_S$ is the maximum possible photosynthesis rate, $E_{PAR}(z)$ and $E^*_{UV}(z)$ are the spectral irradiances of photosynthetically active radiation (PAR, 400-700 nm) and ultraviolet radiation (UV, 280-399 nm) at depth z, and $E_S$ is a parameter measuring the efficiency of the species in using PAR. The asterisk in the ultraviolet irradiance means that spectral UV irradiances are weighted with a biological action spectrum. This is called the E model for photosynthesis (because of its use of irradiances E, instead of fluences H).

*Stellar emission model* The emission spectrum of Alpha Centauri B was obtained using the computer code Spectrum, a stellar spectral synthesis code [4]. The atmospheric model of Kurucz was used [5], with effective temperature $T_{eff}$ = 5316 K, metallicity [Fe/H] = 0,25; log g = 4,44 and microturbulence xi = 1,28 km/s.

It was then assumed a transparent medium between Alpha Centauri B and the top of planetary atmospheres, obtaining the corresponding solar (stellar) constant through:

$$E_{top} = \left(\frac{r}{R}\right)^2 E_{star}, \quad (4)$$

where r is the radius of Alpha Centauri B, R is its distance to the top of the planetary atmosphere, and $E_{star}$ and $E_{top}$ are the irradiances emitted by the star and the one at the top of planet's atmosphere, respectively.

*Climate model* Temperatures $T_S$ at planetary surfaces were calculated using the zero-dimensional climatic model:

$$\frac{E_{top}}{4}(1 - A_{in}) = (1 - A_{out})\sigma T_S^4. \quad (5)$$

In above equation $A_{in}$ and $A_{out}$ are the exoplanet atmospheric albedos to incoming and outgoing radiations, and $\sigma$ is the Steffan-Boltzmann constant.

*Ocean optical models* We use a classification of ocean water according to its optical properties, after extensive experimental works of oceanographer N. Jerlov [6]. He measured attenuation coefficients $K(\lambda)$ of the world ocean in many areas, obtaining a robust classification of ocean water bodies: type I (clearest), type II (intermediate) and type III (darkest). As the reported values of $K(\lambda)$ were given at intervals of 25 nm, we made a linear interpolation to obtain the values for every nanometre in the wavelength range used (280-700 nm). These coefficients were then used to obtain ultraviolet and photosynthetically active irradiances $E_{PAR}(z)$ and $E^*_{UV}(z)$ down the water column, needed to calculate photosynthesis rates (eq. 3). We use a biological action spectrum to account for the photosynthesis inhibition of ultraviolet radiation (thus the asterisk in $E^*_{UV}(z)$). We refer the reader to the work by some of us [7] for details on these calculations.

RESULTS AND DISCUSSION

Tables II and III show results of our computations. We see that the photosynthesis model is related practically in a linear way with planetary surface temperature, giving good photosynthesis rates even in boiling oceans, as in the case for planet c1. This is because; having the same atmosphere, hotter planets would be closer to Alpha Centauri B, receiving more PAR. They would also get more UV, but as it is preferentially attenuated in the upper layer of the ocean, the effect of the increase of PAR is dominant. Figures 1 and 2 illustrate this effect.

Table II
Numerical values for two indexes of habitability in exoplanets around Alpha Centauri B

| Planet | $T_S$ [K] | $\Pi/\Pi_{max}$ | ESI |
|---|---|---|---|
| a1 | 291 | 0,55 | 0,92 |
| a5 | 290 | 0,54 | 0,93 |
| a6 | 271 | 0 | 0,87 |
| a8 | 267 | 0 | 0,91 |
| b2 | 336 | 0 | 0,86 |
| b3 | 349 | 0 | 0,74 |
| b4 | 285 | 0,44 | 0,92 |
| b7 | 310 | 0,43 | 0,91 |
| b8 | 364 | 0 | 0,68 |
| c1 | 373 | 0 | 0,77 |
| c3 | 282 | 0,37 | 0,91 |

Table III
Numerical values for average photosynthesis rates in exoplanets around Alpha Centauri B

| Planet | Water type I | Water type II | Water type III |
|---|---|---|---|
| a1 | 0,72 | 0,51 | 0,36 |
| a5 | 0,72 | 0,51 | 0,36 |
| a6 | 0,68 | 0,48 | 0,34 |
| a8 | 0,67 | 0,47 | 0,34 |
| b2 | 0,77 | 0,57 | 0,40 |
| b3 | 0,78 | 0,58 | 0,40 |
| b4 | 0,71 | 0,50 | 0,35 |
| b7 | 0,75 | 0,54 | 0,38 |
| b8 | 0,79 | 0,60 | 0,41 |
| c1 | 0,80 | 0,60 | 0,42 |
| c3 | 0,70 | 0,50 | 0,35 |

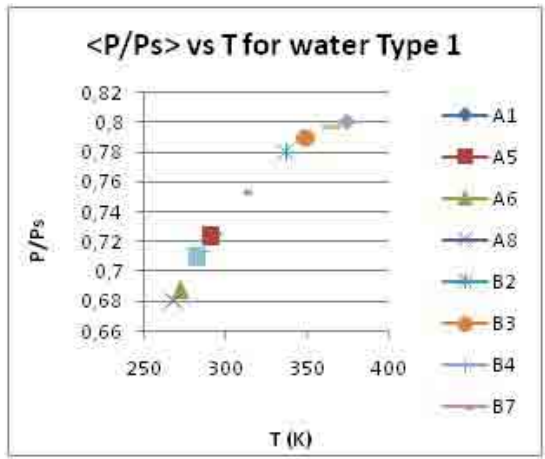

Figure 1: Photosynthesis rates in ocean water type I vs. surface planetary temperature

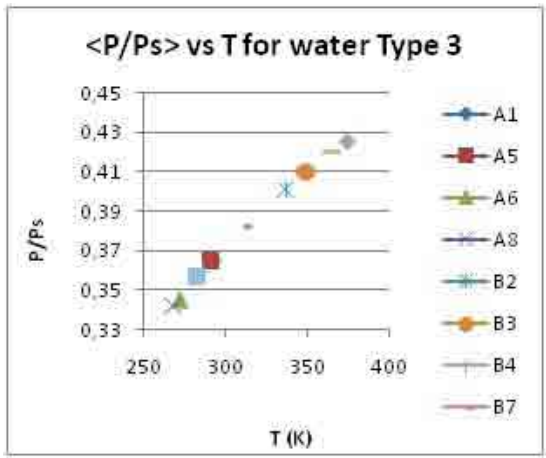

Figure 2: Photosynthesis rates in ocean water type III vs. surface planetary temperature

It can be considered that the Earth Similarity Index (*ESI*) is a good measure of habitability, but this is an index referred to current Earth, which acknowledgeable is not in its maximum potential for life. Actually, the normalized value $\Pi/\Pi_{max}$ for current Earth is estimated to be 0.5 [3]. According to *ESI*, the planets a1, a5, a6, a8, b2, b4, b7 and c3 can host "ordinary Earth-like" life (because its *ESI* falls in the range 0,8 -1,0). The remaining planets (b3, b8 and c1) have *ESI* in the range 0,6 – 0,8; so they could host extremophiles.

The bio-productivity model $\Pi/\Pi_{max}$ is more restrictive, as it by definition considers that life is possible only in the temperature range 0-50 degrees Celsius (provided $T_{opt}$ = 298 K). However, it is noticeable that the five planets giving biological productivity $\Pi/\Pi_{max}$ different from zero coincide with those with highest *ESI* (Table II).

Given the coincident pattern in *ESI* and $\Pi/\Pi_{max}$, it seems clear that the five exoplanets with $\Pi/\Pi_{max}$ different from zero are actually the more suitable for life around Alpha Centauri B.

On another hand, the photosynthesis model clearly fails because it does not consider the surface planetary temperature explicitly.

CONCLUSIONS

Considering the coincident pattern of *ESI* and $\Pi$ bio-model, in this first preliminary assessment we conclude that potential rocky planets a1, a5, b4, b7 and c3 have good potential to harbour photosynthetic life not very different form Earth's, at planetary scale. Planets a1 and a5 seem to have slightly better potential than current Earth, as its normalized $\Pi$ index is greater than 0,5.

The correct inclusion of temperature in the photosynthesis model could yield a third habitability index which might serve as a consistency check of our calculations; this shall be one of our future research directions.